\documentclass{article}
\usepackage{spconf,amsmath,graphicx}
\usepackage{amsfonts}
\usepackage{cite}
\usepackage{booktabs}
\usepackage{multirow}
\usepackage{caption}
\usepackage{url}

\title{Phonetic-aware Speaker Embedding for far-field speaker verification}
\name{Zezhong Jin, Youzhi Tu, Man-Wai Mak \thanks{This work was supported by the RGC of Hong Kong SAR, Grant No. PolyU 15210122.}}
\address{Dept. of Electrical and Electronic Engineering,\\
The Hong Kong Polytechnic University, Hong Kong SAR, China}
\begin{document}
%
\maketitle
\thispagestyle{plain}%
\begin{abstract}
When a speaker verification (SV) system operates far from the sound source, significant challenges arise due to the interference of noise and reverberation. Studies have shown that incorporating phonetic information into speaker embedding can improve the performance of text-independent SV. Inspired by this observation, we propose a joint-training speech recognition and speaker recognition (JTSS) framework to exploit phonetic content for far-field SV. The framework encourages speaker embeddings to preserve phonetic information by matching the frame-based feature maps of a speaker embedding network with wav2vec's vectors. The intuition is that phonetic information can preserve low-level acoustic dynamics with speaker information and thus partly compensate for the degradation due to noise and reverberation. Results show that the proposed framework outperforms the standard speaker embedding on the VOiCES Challenge 2019 evaluation set and the VoxCeleb1 test set. This indicates that leveraging phonetic information under far-field conditions is effective for learning robust speaker representations.
\end{abstract}
\section{Introduction}
\label{intro}
Speaker verification (SV) plays an important role in various fields, such as biometric authentication, e-banking, and access control. Traditional SV models rely on statistical models like Gaussian Mixture Models (GMMs) \cite{reynolds2000speaker} and i-vectors \cite{dehak2010front} to achieve good performance. With the advance in deep learning, deep neural networks, such as TDNNs \cite{waibel2013phoneme}, ResNets \cite{zeinali2019but}, and ECAPA-TDNNs \cite{desplanques2020ecapa}, have been prevailing for speaker embedding. Notably, the ECAPA-TDNN has achieved state-of-the-art performance on various datasets, demonstrating its superiority in speaker verification tasks.

The SV systems mentioned earlier are usually trained on ``clean" utterances and perform well on near-field speech signals. Under far-field conditions, however, due to uncontrollable noise and reverberation, a severe mismatch occurs between the near-field and far-field domains, and these systems suffer greatly \cite{jin2007far}. Developing an SV system that can address the adverse conditions in the far field is essential.

Researchers attempted to address the far-field challenge by modifying the system architecture, exploring adversarial learning techniques, and leveraging advanced data augmentation strategies. For instance, \cite{zhao2021channel} introduced the channel-interdependence enhanced Res2Net (CE-Res2Net) to aggregate speaker information from multi-scale frame-level representations and achieved performance gains on VOiCES Challenge 2019 data. The authors in \cite{yi2020adversarial} used a domain separation network to disentangle and suppress the domain-specific information related to far-field noise and reverberation. In \cite{lin2022robust}, a population-based searching strategy was proposed to optimize the augmentation parameters and greatly boosted far-field SV performance.

On the other hand, studies have shown that text-independent SV systems can be enhanced by incorporating phonetic information into embedding learning. In \cite{liu2018speaker}, the authors adopted a multi-task learning strategy by combining a phonetic classifier with a speaker classifier for speaker embedding and obtained superior performance. The authors of \cite{wang2019usage} investigated the usefulness of phonetic information at the segment level and the frame level. They concluded that although phonetic content at the segment (embedding) level is detrimental to SV performance, using phonetic information at the frame level is beneficial. One possible explanation for the performance improvement in \cite{liu2018speaker, wang2019usage} is that shared spectral dynamics exist at the lower (frame-level) layers, which are useful for speech and speaker recognition. Enriching content information at the frame-level layers also strengthens the information essential for speaker discrimination.
\begin{figure*}[!t]
    \centering
    \includegraphics[]{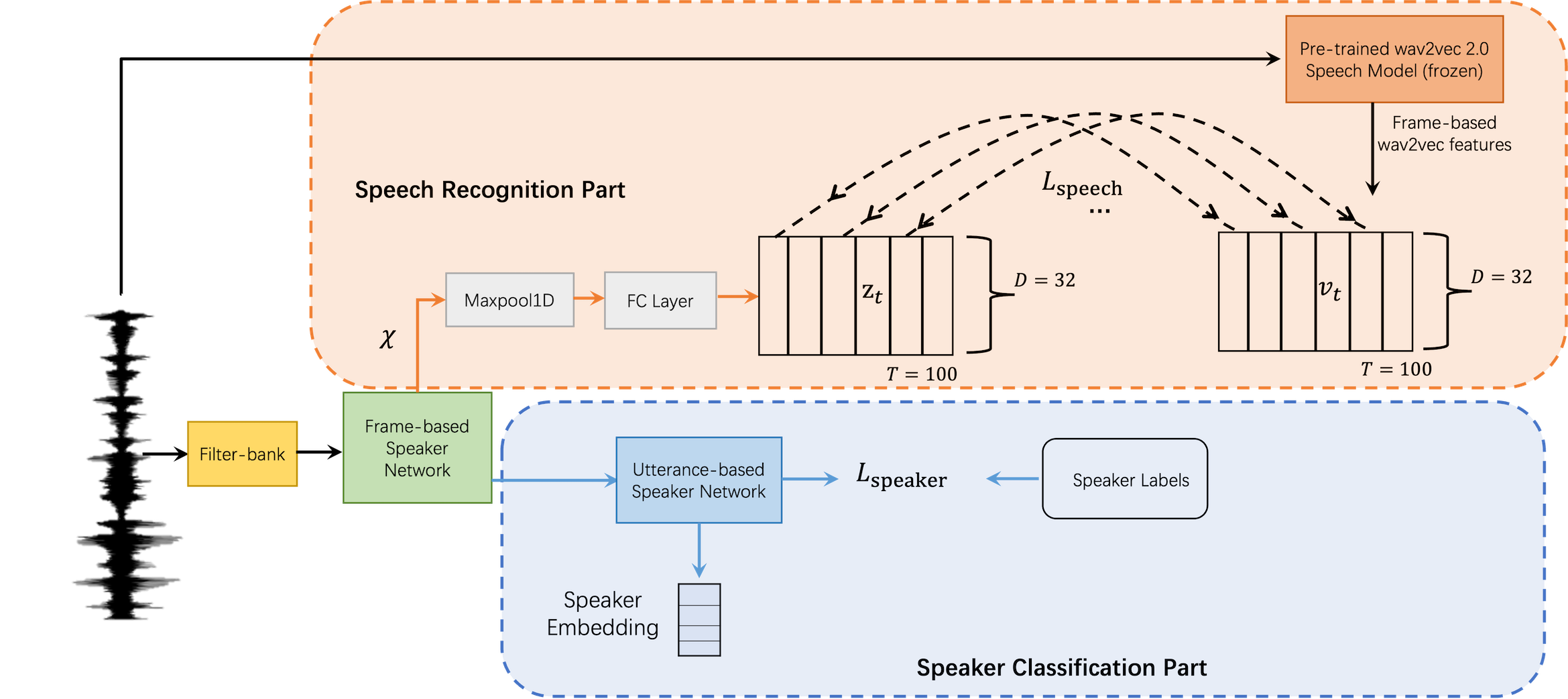}
    \caption{Framework of joint training of speech recognition and speaker classification (JTSS). The utterance-based speaker network in the speaker classification part comprises a pooling layer and a fully connected layer. }
    \label{JTSS}
\end{figure*}

In this paper, we exploit phonetic information in far-field SV. Inspired by the above observations, we propose a framework that can jointly train a model to perform speech recognition and speaker verification tasks. The framework comprises
a speech recognition component for phonetic information extraction and a speaker identification component for enforcing the segment-level layers to produce speaker discriminative vectors. We refer to the framework as {\bf j}oint-{\bf t}raining {\bf s}peech and {\bf s}peaker recognition (JTSS). Unlike \cite{liu2018speaker,wang2019usage}, phonetic labels are not required in JTSS.  Instead, we use a pre-trained wav2vec 2.0 model to extract phonetic content in an unsupervised way. This strategy greatly saves the effort to transcribe the speech files in speaker recognition corpora. The rationale behind JTSS is that although noise and reverberation can blur speaker information in speech signals, the phonetic information extracted from wav2vec 2.0 assists in preserving the underlying acoustic dynamics shared by the speaker identity. Therefore, the degradation due to the far-field conditions can be compensated to a certain extent. Our main contributions are as follows:
\begin{enumerate}
    \item We proposed a phonetic-aware JTSS framework, which improves the robustness of far-field SV by exploiting phonetic information.
    \item We incorporated a pre-trained wav2vec 2.0 model in the speech recognition part, eliminating the need for manually transcribing speaker verification datasets.
\end{enumerate}
The rest of the paper is organized as follows. Section 2 introduces the wav2vec 2.0 and details the JTSS framework. Section 3 presents the experimental settings, and Section 4 shows the results and analyses. We draw a conclusion in Section 5.

\section{Methodology}
\label{method}
This section introduces the JTSS framework and its two components: the speech recognition part and the speaker classification part.
\vspace{0.5cm}
\subsection{Speech Recognition Model}
In Fig.~\ref{JTSS}, we utilize a wav2vec 2.0 \cite{baevski2020wav2vec} network fine-tunned by CTC loss \cite{graves2006connectionist} as the speech model. Wav2vec 2.0 is a self-supervised learning framework that leverages a large amount of unlabeled data to learn speech representations. It takes in a waveform and produces context representations through a stack of CNN layers and transformer layers. Through contrastive learning, the model is able to extract compact and meaningful speech representations that can be used for downstream speech tasks. Recently, the pre-trained wav2vec 2.0 model has gained popularity as a front-end feature extractor in various speech applications.
\subsection{JTSS Framework}
As shown in Fig.~\ref{JTSS}, the speech recognition component and the speaker classification component share the frame-level layers (green block). The representations outputted from an intermediate frame-level layer are fed into the speech recognition part. We denote these representations as $\mathcal{X} = \{ \bold{x}_t \in \mathbb{R}^D; t =1,\dots,T \}$, where $\bold{x}_t$ is a $D$-dimensional vector at the $t$-th frame. For the speaker classification part, the feature maps produced from the last frame-level layer are processed by a pooling layer and a fully connected (FC) layer to derive an utterance-level embedding $\bold{e}$. The AAMSoftmax \cite{xiang2019margin} loss is employed as the loss function ($L_\mathrm{speaker}$ in Fig~\ref{JTSS}).

For the speech recognition part, the waveform is fed into the speech model and we obtain a sequence of $T$ frames $\mathcal{V} = \{ \bold{v}_t \in \mathbb{R}^{\widetilde{D}}; t =1,\dots, T\}$, where ${\widetilde{D}}$ is the dimension of speech vectors. A max-pooling layer is applied to $\mathcal{X}$ to ensure that the resulting $\mathcal{Z}=\{\bold{z}_t \in \mathbb{R}^{\widetilde{D}} ; t=1,\dots,T \}$ has the same length as $\mathcal{V}$.  We compute the speech loss as the cosine similarity between $\mathcal{Z}$ and $\mathcal{V}$:
\begin{equation}
    L_\mathrm{{speech}} = 1 -\frac{1}{T}\sum_{t=1}^T\mathrm{cos}(\bold{z}_t,\bold{v}_t)   .
\end{equation}
Then, we average the $L_\mathrm{speech}$ across the utterances in a mini-batch.
By making $\mathcal{Z}$ close to $\mathcal{V}$, we enable the frame-level layers of the speaker encoder to preserve useful phonetic information. Because phonetic information contains speaker-dependent acoustic dynamics, maintaining phonetic information at the frame level would also preserve speaker information in the embedding network. As will be demonstrated in Section 4.1, this speaker information preservation helps compensate for the performance degradation caused by far-field environments.

The total loss is defined as follows:
\begin{equation}
    L_\mathrm{{total}} = L_\mathrm{{speaker}} + \lambda L_\mathrm{{speech}} ,%
    \label{eq2}
\end{equation}
where $L_\mathrm{{speaker}}$ is the AAMSoftmax loss defined in \cite{xiang2019margin} and $\lambda$ is a hyperparameter that controls the contribution of phonetic information. During training, we freeze the parameters of the speech model.
\section{Experimental Setup}
\label{experiment set}
\subsection{Datasets and Data Preparation}
The training data comprise the VoxCeleb1 development set and the VoxCeleb2 development set \cite{nagrani2017voxceleb} \cite{chung2018voxceleb2}, which consist of a total of 7,205 speakers. Voice activity detection (VAD) was not used. We followed the data augmentation strategy in Kaldi's recipes \cite{povey2011kaldi}. We added noise, music, and babble to the training data using MUSAN \cite{snyder2015musan} and created reverberated speech data based on RIR \cite{jeub2009binaural}. For evaluation, we used the VOiCES Challenge 2019 evaluation (VOiCES19-eval) dataset \cite{nandwana2019voices}. The Voxceleb1 test Original (Vox-O), which comprises 40 speakers, was also used as the evaluation set. 

\subsection{Network Training}
We used the standard x-vector \cite{snyder2018x} and ECAPA-TDNN \cite{desplanques2020ecapa} as our backbones. The channel size of ECAPA-TDNN is 512. The dimension of speaker embeddings is 192 for ECAPA-TDNN and 512 for x-vector, respectively. For the speech model, we used the wav2vec 2.0 model fine-tuned on the LibriSpeech dataset \cite{panayotov2015librispeech}.\footnote{\url{https://huggingface.co/facebook/wav2vec2-base-960h}} The output of the wav2vec 2.0 was obtained from the projection layer of the fine-tuned model. The frame-level representation from the lowest-level TDNN of the x-vector network and the ECAPA-TDNN were used as the input to the speech recognition part.

For ECAPA-TDNN, we extracted 80-dimensional filter-bank (Fbank) features from 16 kHz audio signals using a 25ms window with a 10ms frameshift. For the x-vector network, we extracted 40-dimensional Fbank features. Each training segment in the mini-batch has a duration of 2 seconds. The batch size was set to 100 for ECAPA-TDNN and 50 for x-vector, respectively. We used an Adam optimizer with an initial learning rate of 0.001 and employed a step learning rate scheduler. The total number of epochs is 80. For the AAMSoftmax loss function, the margin is 0.2 and the scale is 30.

\subsection{Performance Evaluation}
We used a cosine backend in all experiments. When performing evaluation on the Vox-O test set, we followed the setting in \cite{desplanques2020ecapa} to apply the AS-norm \cite{matejka2017analysis} on the scores. The performance metrics include equal error rate (EER) and minimum detection cost function (minDCF) with $P_\mathrm{target} =$ 0.01.
\section{Results and Analyses}
\label{Results and Analyses}
\renewcommand\arraystretch{1.3}    
\begin{table}[t]
        \normalsize
        \centering
        \captionsetup{skip=10pt}
        \caption{Comparison of our method (JTSS) with other methods on the VOiCES19-eval dataset. The best results for each model are highlighted in bold.
}
\label{table1}
        {\setlength{\tabcolsep}{3pt}
        \begin{tabular}{ccccc}
            \toprule[0.8pt]
            System &Speaker Embedding &EER (\%)&minDCF \\
            \midrule[0.8pt]
            \cite{zhao2021channel} &CE-Resnet&5.72&0.423 \\
            \cite{novoselov2019stc} &x-vector& 8.55 &0.552 \\
            \cite{dowerah2023joint} &ECAPA-TDNN& 5.90 & \_ \\
            \midrule[0.8pt]
            Baseline 1& x-vector&7.81&0.598\\
            JTSS (Proposed) &x-vector&6.85 & 0.483\\
            Baseline 2&ECAPA-TDNN&5.79&0.428\\
            JTSS (Proposed)&ECAPA-TDNN& \textbf{5.13} & \textbf{0.374}\\
            \bottomrule[0.8pt]
        \end{tabular}}
    \end{table}
    \begin{table}[t]
        \normalsize
        \centering
        \caption{Comparison of our method (JTSS) with our
    baseline on the clean and noisy Vox-O datasets.}
        \label{table2}
        {\setlength{\tabcolsep}{1pt}
        \begin{tabular}{cccccc}
            \toprule[0.8pt]
            System &Speaker Embedding&\multicolumn{2}{c}{Vox-O (clean)}&\multicolumn{2}{c}{Vox-O (noise)}\\
            &&EER&minDCF &EER&minDCF\\
            \midrule[0.8pt]
            Baseline 1&x-vector&2.23 & 0.219&5.82&0.470\\
            JTSS&x-vector&2.16&0.205&5.14&0.446\\
            Baseline 2&ECAPA-TDNN&1.19&0.165&3.95&0.380 \\
            JTSS&ECAPA-TDNN&\textbf{1.10}&\textbf{0.136}&\textbf{3.31}&\textbf{0.311} \\
            \bottomrule[0.8pt]
        \end{tabular}}
    \end{table}
    
\begin{table}[t]
        \normalsize
        \centering
        \caption{Impact of preserving phonetic information at different frame-level layers of an ECAPA-TDNN.}
        \label{table3}
        {\setlength{\tabcolsep}{4pt}
        \begin{tabular}{cccc}
            \toprule[0.8pt]
           Frame-level Layer&Network Block&EER (\%)&minDCF \\
            \midrule[0.8pt]
            Layer 4&TDNN&5.28 & 0.392 \\
            Layer 3&SE-Res2Block& 5.54&0.379 \\
            Layer 2&SE-Res2Block&5.33 &0.385 \\
            Layer 1&SE-Res2Block&5.27& 0.390 \\
            Layer 0&TDNN&\textbf{5.13}&\textbf{0.374} \\
            \bottomrule[0.8pt]
        \end{tabular}}
    \end{table}
\begin{table}[t]
        \normalsize
        \centering
        \caption{Impact of $\lambda$ (in eq. \ref{eq2}) on the proposed framework. The best results are highlighted in bold.}
        \label{table4}
        {\setlength{\tabcolsep}{8pt}
        \begin{tabular}{cccc}
            \toprule[0.8pt]
            Speaker embedding&$\lambda$&EER (\%)&minDCF \\
            \midrule[0.8pt]
            \multirow{6}{*}{ECAPA-TDNN}&0.001&5.69 &0.410 \\
            &0.004&5.67&0.418\\
            &0.01&5.59&0.401\\
            &0.1&\textbf{5.13} & \textbf{0.374} \\
            &0.4 &5.85&0.448\\
            \bottomrule[0.8pt]
        \end{tabular}}
    \end{table}
We report the performance of JTSS in this section. The comparison with conventional speaker embeddings is detailed.
\vspace{0.3cm}
\subsection{Main Results}
Table \ref{table1} presents the results of various systems on the VOiCES19-eval dataset. We observe that our baselines achieve superior or comparable performance to existing systems. From Table \ref{table1}, it is evident that JTSS outperforms the baselines for both x-vector and ECAPA-TDNN. Specifically, for ECAPA-TDNN, our proposed method reduces the EER by  12.9\% and minDCF by 14.4\%. For x-vector, our method achieves a reduction of 14.1\% and 23.8\% on EER and minDCF, respectively. This observation demonstrates the effectiveness of JTSS in leveraging phonetic information for far-field SV.

To verify that JTSS can partially compensate for the degradation due to adverse conditions in the far field, we investigated the performance of JTSS on the clean and noisy Vox-O datasets. The ``clean" set refers to the standard Vox-O test data, and the noisy Vox-O set was created by randomly adding noise and reverberation to the standard (clean) Vox-O data, following the data augmentation strategy in Section 3.2.  Table \ref{table2} shows the results of JTSS and the baseline models. From Table \ref{table2}, we observe that JTSS outperforms the baselines on both clean and noisy Vox-O sets. On the clean Vox-O, JTSS achieves a slight improvement over the baseline. This confirms the conclusion in \cite{liu2018speaker,wang2019usage} that using phonetic content can benefit text-independent SV. On the noisy Vox-O set, we see substantial performance degradation compared with the clean counterpart. Nevertheless, JTSS obtains remarkably greater performance gains over the baseline systems. This observation verifies our motivation that incorporating phonetic information into the speaker embedding system can improve SV performance, particularly in far-field environments with noise and reverberation.
\subsection{Ablation Study}
Table \ref{table3} shows the impact of feeding different frame-level representations to the speech recognition part on SV performance. In Table \ref{table3}, Layer 0 and Layer 4 correspond to the initial and final TDNN layers of the ECAPA-TDNN, respectively. The remaining three layers correspond to the three SE-Res2Blocks, respectively. 

Table \ref{table3} shows that the performance improvement of JTSS becomes more prominent when we feed features from lower layers into the speech recognition part. Specifically, when we input features from the initial TDNN layer (Layer 0) into the speech recognition part, we obtained the best result with an EER of 5.13\%. However, performance gradually drops when we preserve phonetic information at the upper layers (with higher-level representations). This result is reasonable because the lower-level feature maps contain more speaker and content information that is entangled together. By contrast, the representations at upper layers are more speaker-specific. Therefore, it is preferable to exploit phonetic information at lower layers. This is also the reason for using the bottom layer for phonetic information extraction in Section 4.1.

We also investigated the effect of $\lambda$ in Eq. \ref{eq2} on JTSS. The results are shown in Table \ref{table4}. We observe that the best performance is achieved when $\lambda=0.1$, with an EER of 5.28\%. As $\lambda$ increases, the performance of JTSS gradually deteriorates. When $\lambda$ was set to 0.4, the EER of the JTSS system is higher than that Baseline 2 in Table \ref{table1}. The above observations suggest that excessive phonetic information can cause the speaker embedding network to focus on the content details, neglecting the speaker information and leading to performance degardation.
\section{Conclusions}
\label{conclusions}
In this paper, we propose a joint training framework (JTSS) for speech recognition and speaker verification tasks to improve far-field SV performance. By using a pre-trained speech recognition model, we incorporate the phonetic information into the conventional speaker encoders. Also, we eliminate the reliance on transcriptions for the speech recognition task. Experimental results demonstrated that leveraging phonetic information can improve the performance of far-field speaker verification.
\small
\bibliographystyle{IEEEbib}
\bibliography{strings,refs}

\end{document}